\documentclass[conference,a4]{IEEEtran}
\IEEEoverridecommandlockouts

\usepackage{cite}
\usepackage{amsmath,amssymb,amsfonts}
\usepackage{algorithmic}
\usepackage{graphicx}
\usepackage{textcomp}
\def\BibTeX{{\rm B\kern-.05em{\sc i\kern-.025em b}\kern-.08em
    T\kern-.1667em\lower.7ex\hbox{E}\kern-.125emX}}

\usepackage[T1]{fontenc}

\usepackage{caption}
\usepackage{subcaption}
\usepackage{amssymb}
\usepackage{wrapfig}

\usepackage{longtable}
\usepackage[table]{xcolor}
\usepackage{eqparbox}
\usepackage{float}
\usepackage{lipsum}
\usepackage{multirow}
\usepackage{scalerel}
\usepackage{hyperref}
\usepackage{amsmath}
\usepackage{mathtools}
\usepackage{microtype}
\usepackage{enumitem}
\usepackage[capitalize]{cleveref}
\Crefname{Chapter}{Chap.}{Chaps.}
\Crefname{Section}{Sec.}{Secs.}
\Crefname{Figure}{Fig.}{Figs.}

\usepackage{glossaries}
\glsdisablehyper
\newacronym{asr}{ASR}{automatic speech recognition}
\newacronym{lm}{LM}{language model}
\newacronym{ilm}{ILM}{internal language model}
\newacronym{tts}{TTS}{text-to-speech}
\newacronym{seq2seq}{seq2seq}{sequence-to-sequence}
\newacronym{ctc}{CTC}{connectionist temporal classification}
\newacronym{aed}{AED}{attention-encoder-decoder}
\newacronym{fh}{FH}{factored hybrid}
\newacronym{trans}{mRNN-T}{strictly monotonic recurrent neural network transducer}
\newacronym{bpe}{BPE}{byte-pair encoding}
\newacronym{lstm}{LSTM}{long short-term memory}
\newacronym{oov}{OOV}{out-of-vocabulary}
\newacronym{mtg}{MTG}{MTGJSON}
\newacronym{lbs}{LBS}{LibriSpeech}
\newacronym{wer}{WER}{word error rate}
\newacronym{ppl}{PPL}{perplexity}

\def\am{$\theta_{\scaleto{\text{AM }}{4pt}}$}
\def\lm{$\theta_{\scaleto{\text{LM}}{4pt}}$}

\newcommand{\argmax}{\operatornamewithlimits{argmax}}
\newcommand\numberthis{\addtocounter{equation}{1}\tag{\theequation}}
\def\lbs{Real-LBS}

\def\med{MEDLINE}

\usepackage{xcolor}

\def\cameraready{1}

\begin{document}
	
\title{Analysis of Domain Shift across ASR Architectures via TTS-Enabled Separation of Target Domain and Acoustic Conditions\\
	\thanks{$^\star$ Denotes equal contribution}
}


\author{\IEEEauthorblockN{Tina Raissi$^\star$}
	\IEEEauthorblockA{\textit{Machine Learning and} \\
		\textit{Human Language Technology Group}\\
		\textit{RWTH Aachen University}\\
		Aachen, Germany \\
		raissi@ml.rwth-aachen.de}
	\and
	\IEEEauthorblockN{Nick Rossenbach$^\star$}
	\IEEEauthorblockA{\textit{Machine Learning and} \\
		\textit{Human Language Technology Group}\\
		\textit{RWTH Aachen University}\\
		\textit{AppTek GmbH}\\
		Aachen, Germany \\
		rossenbach@ml.rwth-aachen.de}
	\and
	\IEEEauthorblockN{Ralf Schl\"uter}
	\IEEEauthorblockA{\textit{Machine Learning and} \\
		\textit{Human Language Technology Group}\\
		\textit{RWTH Aachen University}\\
		\textit{AppTek GmbH}\\
		Aachen, Germany \\
		schlueter@ml.rwth-aachen.de}
}

\maketitle

\begin{abstract}
	
We analyze \gls{asr} modeling choices under domain mismatch, comparing classic modular and novel \gls{seq2seq} architectures.\ Across the different ASR architectures, we examine a spectrum of modeling choices, including label units, context length, and topology. To isolate language domain effects from acoustic variation, we synthesize target domain audio using a text-to-speech system trained on LibriSpeech. We incorporate target domain n-gram and neural language models for domain adaptation without retraining the acoustic model. To our knowledge, this is the first controlled comparison of optimized ASR systems across state-of-the-art architectures under domain shift, offering insights into their generalization.\ The results show that, under domain shift, rather than the decoder architecture choice or the distinction between classic modular and novel \gls{seq2seq} models, it is specific modeling choices that influence performance.
\end{abstract}


\glsresetall

\section{Introduction}
The research community has shown increasing interest in \gls{seq2seq} approaches that integrate the optimization of both acoustic and language models in a unified framework, commonly referred to as end-to-end.\ 
Currently, the predominant landscape of \gls{asr} models consists of time synchronous finite state machines with various label topologies and different frame level posteriors definition~\cite{graves2006connectionist,graves2012sequence,tripathi2019monotonic}, along with the label synchronous models equipped with attention mechanism~\cite{chorowski2015attention}.\ 
A closer comparative analysis reveals however a continuity between the classic hybrid neural network hidden Markov~(NN-HMM) models~\cite{Bourlard+Morgan:1993} and the recent approaches~\cite{prabhavalkar2023end}.\ 
While end-to-end systems are often associated with greater simplicity and improved performance, conducting a rigorous comparison with the classic approaches can present several challenges.\ 
Among these are differences in the amount of training data and optimization steps, neural encoder architectures, and the extent with which external resources such as language models or pronunciation lexica are utilized.\ Moreover, while models are often identified by their associated frameworks of specific training paradigms and resource dependencies, it is important to decouple the conceptual definition of a model from its particular implementation. This distinction enables a more principled and fair comparison across approaches, independent of auxiliary design choices.\ Principled comparisons that take these factors into account indicate that the evaluation of performance superiority remains inconclusive~\cite{rouhe2023principled,raissiinvestigating, gimeno2024comparison}.\

One research area where the strict divide between end-to-end and hybrid HMM systems becomes particularly evident is domain mismatch during recognition.\ This binary comparison is commonly supported by differences in the formulation of the Bayesian decision rule.\ In the classic approach, the joint probability of the input speech features and the output word sequence is factorized into separate language and acoustic models, with further factorization of the latter into an alignment state transition and an optional pronunciation model.\ In contrast, the word sequence posterior in a \gls{seq2seq} framework is defined to implicitly include all the mentioned factors in a single model.\ The end-to-end training of these models leverages large amounts of paired audio and text data, and the flexibility of the blank-augmented alignment topologies, as well as larger label units.\ 

However, since the unpaired text data is usually available in much larger quantities, incorporating an external language model trained on a larger and potentially more domain relevant data can enhance the system's robustness to the domain shift.\ 
One of the main issues in incorporating an external language model that includes statistics on the text data with mismatched domain is the presence of an \gls{ilm}~\cite{VarianiRybachAllauzen+20}.\ This is implicitly learned from the audio transcriptions not only when the optimized label posterior is conditioned on a label history, but also in models that assume frame-level conditional independence~\cite{zhaoregarding}.\ 
Existing methods suppress or adapt the \gls{ilm} via additional training with complex pipelines, via specific joint / factorized modeling, or estimate and subtract it during inference~\cite{Michel+Schlueter+2020:lm_integration,adapt1, adapt2,adapt4,adapt5,factoredrnnt1,factoredrnnt2,factoredrnnt3,factoredrnnt4,zeineldeen2021:ilm,Zhou+Zheng+:2022}.\

\subsection{Contribution}

In this work, we show that domain shift behavior and adaptability cannot be fully explained by the simple division between novel \gls{seq2seq} models and classic hybrid HMM approaches. Our goal is not to exhaustively enumerate all factors that influence this behavior, but rather to provide an initial case study through selected examples. To this end, we conduct a principled comparison of \gls{asr} model performance under domain shift during inference.\ 
We consider a selection of the most popular models and separate the model definitions and training configurations from the conventional implementation practices used in current frameworks in the community.\ 
We keep the training and inference conditions comparable across architectures, where possible.\ 
Through a set of targeted comparisons, we showcase modeling choices that contribute to improved \gls{asr} robustness under domain shift.\

Our primary focus is to study the effect of language model domain shift, a task that requires careful handling, as it is crucial to minimize interference from acoustic conditions and variations.\ To address this delicate aspect of the evaluation we adopt the following strategy: while all \gls{asr} architectures are trained on real audio, we employ a \gls{tts} system to generate test data that maintains similar acoustic conditions but belongs to a different language model domain.\ 
Recent works have shown that synthetically generated data does perform reasonably well for domain adaptation purposes \cite{dhamyal24_syndata4genai} and \gls{asr} training without adding real data \cite{rossenbach24_syndata4genai}. 
Utilizing recent advances in temperature controlled generative modeling, we configure the \gls{tts} system to generate synthetic data which is recognized with error rates matching those of real data.\  
Using this novel setting, we generate two synthetic test datasets with similar acoustic conditions in a biomedical and a trading card game domain.\ 
We implement a selection of popular ASR architectures, deliberately diverging from their canonical configurations in some cases.\ 
Specifically, by decoupling the common architectural design from core modeling choices and assumptions, we build a first-order label context phoneme transducer\cite{zhou:phoneme-transducer:2021} and two fully neural factored hybrid HMMs with untied diphone and triphone labels~\cite{raissi2023competitive}, i.e., without decision trees \cite{young1994tree}. \ In addition to a phoneme based \gls{ctc}, we also include two representatives of time- and label- synchronous model families with \gls{bpe} label units and full label history, namely a transducer and an \gls{aed} model.
We estimate a 4-gram \gls{lm} and train a \gls{lstm} \gls{lm} to show the domain adaptation capability of each different \gls{asr} system.\ More details on the experimental design follow in section \ref{sec:expdes}.
To our knowledge, there are no prior comparative studies across different \gls{asr} architectures with the described conditions.\ 

\section{Experimental Design}
\label{sec:expdes}

In order to study the effect of the language model independently from the acoustic condition variations, we first train \gls{asr} models and a \gls{tts} model on real data, i.e., on the \gls{lbs} corpus \cite{povey2015librispeech}.\ 
We then generate synthetic test data using text from a different domain.\ To ensure a fair comparison, we adhere to the following criteria: (1) our \gls{asr} models are trained under conditions that result in similar performance ranges on the real \gls{lbs} dev and test sets, ensuring that all models perform comparably on real data, (2) We select a \gls{tts} system capable of generating synthetic LBS development data on which our  \gls{asr} systems achieve accuracy comparable to their performance on real data.
Where possible, the training and inference conditions are kept consistent and comparable across architectures.\ The reader is encouraged to interpret our results within the following four-fold comparative scenario:

\begin{enumerate}
	\item \textbf{First-order label context}: \gls{trans} and \gls{fh} models with phoneme label units
	\item \textbf{Two \gls{fh} models:} one with only left context (diphone) and one with additional right context (triphone)
	\item \textbf{Two \gls{trans}  models:} one phoneme-based with first-order label context and one with full label history and \gls{bpe} label units
	\item A \textbf{phoneme based \gls{ctc}} and a \textbf{BPE based \gls{aed}}, for completeness
 \end{enumerate}

It is commonly believed that hybrid HMM models are more robust to domain shift than transducers. To evaluate this claim, in our initial comparison we examine a transducer and a \gls{fh} under fair and controlled conditions.\
The second comparison highlights the significance of right-context modeling, which is consistently handled only in the hybrid HMM framework.\ This stems not only from the generative starting point in the hybrid HMM, but also the blank-free label topology which enables a consistent HCLG-like search space representation~\cite{mohri2002weighted,nolden2017phd}.
Finally, we contrast the phoneme-based \gls{ctc} model with those from the first case study due to its label independence assumption, and include the standard \gls{bpe} based \gls{aed} model as a representative of fully end-to-end \gls{asr} systems.
We use two different \gls{lm}s to highlight the domain adaptability of each model when using weaker count-based and full context neural linguistic statistics.\ The integration of the (external) \gls{lm} takes advantage of the recent findings on the importance of the subtraction of the \gls{ilm}.\ While the factored hybrid models use context-dependent state priors during inference, we apply \gls{ilm} subtraction to all \gls{seq2seq} models.\

\section{Overview of Models}
The Bayes decision rule for the ASR task maximizes the a-posteriori probability of a word sequence $W$ given an input acoustic feature sequence $X$~\cite{bayes1763lii}.\ 
The sequence-level class posterior probability in \cref{eq:bayes1} is the usual discriminative starting point in seq2seq approaches, for a given joint acoustic and language model set of parameters $\theta$.\ 
In the classic generative approach, the optimization can be done for separate acoustic and language models  parameters \am and \lm, respectively.\ 
This follows the equivalent formulation in \cref{eq:bayes2} obtained via Bayes identity.
\begin{equation}\footnotesize
	\label{eq:bayes1} 
	X \rightarrow\tilde{W}(X) = \underset{W}{\argmax} \left\lbrace P_{_{\theta}}(W | X)  \right\rbrace \phantom{0000000000} \vspace{-0.3cm}
\end{equation}
\begin{equation}
	\label{eq:bayes2}\footnotesize
	\phantom{X \rightarrow\tilde{W}(X)}= \underset{W}{\argmax} \left\lbrace P_{_{\theta_{\scaleto{\text{AM}}{3pt}}}} (X | W) \cdot P_{_{\theta_{\scaleto{\text{LM}}{3pt}}}}(W) \right\rbrace 
\end{equation}
We study the effect of the domain shift on \gls{asr} accuracy across one label-synchronous model and three different time-synchronous acoustic models, each varying in label unit and label context order.\ 
We denote by $h_1^T$ the acoustic encoder output which transforms $X$ into high-level representations with subsampling.\ 
Each output label sequence $\phi$ of length $M$ corresponding to $W$ consists of specific model label units. It is unfolded in time via the marginalization over alignment sequences when required by the modeling approach.\ 

We define a generic function $a_n(\cdot)$ to identify the label unit identity within the word at position $n$.\ 
We overload this function in \cref{subsec:fh,subsec:ctc-trans,subsec:aed} to match each model definition, accordingly.\ 
Introducing this function is crucial for achieving a uniform notation. Moreover, it addresses the need to decouple alignment state identity from arbitrary lexical label choices.

\subsection{Factored Hybrid HMM}
\label{subsec:fh}
We consider \gls{fh} HMM as a generative acoustic model with parameter \am from \cref{eq:bayes2}.\ For a phoneme sequence $\phi$ and hidden Markov alignment state sequences $s_1^T$, we define $a_{n_{s_t}}$ to be a mapping function that accepts as input the aligned state at time frame $t$ within a phoneme of the word at position $n$ and returns its label.\ In triphone \gls{fh}, by incorporating the identity of the right and left phoneme label contexts at each time frame, the joint probability defined in \cref{eq:fh} can be factorized into separate label posterior factors~\cite{bourlard1992cdnn,raissi2020fh}.\ By omitting the right phoneme, the label context span can be reduced, yielding a diphone model.\ 
The decision rule described in \cref{eq:fh} employs a state transition model with only label loop and forward, and a frame-level label prior with exponents $\alpha$, and $\beta$, respectively. We combine the language model~(LM) with exponent  $\lambda$ according to \cref{eq:bayes2}.\

\begingroup
\addtolength{\jot}{-1em}
\begin{align*} 
	\label{eq:fh}  \numberthis  
	\underset{W}{\argmax} \hspace{-0.5mm} & \left. \Bigg\{ \hspace{-0.5mm}P^{\lambda}_{\text{LM}}(W) \right. \\ 
	& \left. \underset{\scaleto{s_1^T:\phi_1^M:W}{8pt}}{\max} \prod_{t=1}^T \hspace{-1mm} \frac{P(a_{n_{s_t}+1}, a_{n_{s_t}}, a_{n_{s_t}- 1})| h_t)}{P^{\alpha}_{\text{Prior}}(a_{n_{s_t}+1}, a_{n_{s_t}}, a_{n_{s_t}-1})} P^{\beta}(s_t| s_{t-1})  \right\rbrace \nonumber
\end{align*} 
\endgroup
\subsection{CTC and Transducer}
\label{subsec:ctc-trans}

For the direct discriminative approach of \cref{eq:bayes1}, we explore \gls{ctc}~\cite{graves2006connectionist} and two \gls{trans}~\cite{tripathi2019monotonic,Zhou+Michel+:2022}, and the \gls{aed}~\cite{chorowski2015attention}, described later in \cref{subsec:aed}.\ 
The two \gls{trans} models differ in label unit and context length as follows: one is phoneme-based and restricted to a single output label context, while the second uses \gls{bpe}~\cite{sennrich-etal-2016-neural} and supports unlimited context, i.e., full output label sequence context-dependency.\ 
The general formulation with infinite context is shown in \cref{eq:trans-decode}, with $y_1^T$ denoting the blank augmented alignment sequence corresponding to the output sequence $\phi$, which here stands for a \gls{bpe} or a phoneme sequence.\ 
We overload the $a_{y_t}$ function to accept the alignment state $y_t$ at time frame $t$ and return the most recent emitted output label.\ 
For both \gls{trans} models, in order to combine an external LM we divide the label posterior by a the \gls{ilm} estimated based on the label context order.\ 
Dropping the label context-dependency $a_1^{y_{t-1}}$ results in the decision rule for \gls{ctc}, with the difference that, similar to \gls{fh}, a label prior $P^\alpha_{PR}(y_t)$ is used instead of the \gls{ilm}.\ 
\begin{equation}\footnotesize
	\underset{W}{\argmax} \hspace{1mm}  \left\lbrace P^{\lambda}_{\text{LM}}(W) \cdot \underset{\scaleto{y_1^T:\phi_1^M:W}{8pt}}{\max} \prod_{t=1}^T \frac{P(y_t | a_1^{y_{t-1}}, h_1^T )}{P^{\alpha}_{\text{ILM}}(y_t | a_1^{y_{t-1}})}  \right\rbrace \label{eq:trans-decode}
\end{equation} 

\subsection{Attention Encoder-Decoder}
\label{subsec:aed}
We consider a \gls{bpe}-based \gls{aed} as the label synchronous discriminative model with global attention mechanism.\ 
With the absence of an alignment sequence, the function $a_m$ returns the \gls{bpe} label at position $m$.\ 
The decision rule includes also a sequence length normalization term with the exponent $\delta$.

\begin{equation} \footnotesize
	\label{eq:attention-decode}
	\hspace{-0.3cm} \underset{\{M,a_1^M:W\}}{\argmax} \hspace{1mm}  \left\lbrace  \frac{1}{M^{\delta}} \prod_{m=1}^M  P^{\lambda}_{\text{LM}}(a_m | a_1^{m-1}) \frac{P(a_{m} | a_{1}^{m-1}, h_1^T)}{P^{\alpha}_{\text{\gls{ilm}}}(a_m | a_1^{m-1}) }  \right\rbrace 
\end{equation} 

\section{Speech Synthesis}
\label{subsec:TTS}
We use Glow-TTS \cite{kim_glow-tts_2020} as the \gls{tts} architecture to generate the synthetic data.\ 
Given an invertible decoder function $f$, an audio feature sequence $x_1^T$  is produced based on a Gaussian distributed latent variable $z_1^T$ as follows:
\begin{equation} \label{eq:tts}
	{x}_1^T = f^{-1}_{\theta}({z}_1^T) \text{, where }
	{z}_1^T \sim \mathcal{N}(\mu_{\theta}({h'}_1^T),\tau\mathbf{1})
\end{equation}
The latent variable $z$ is sampled based on a mean determined by a mean predictor layer $\mu_\theta$ on top of an up-sampled text encoder network output ${h'}_1^T$. 
A unit vector scaled by a temperature factor $\tau$ is used as variance. 
The temperature factor allows to control the variability of the produced audio features. 
With this factor we can control the word error rate which the different \gls{asr} systems will have on generated data. We use Griffin \& Lim \cite{DBLP:conf/icassp/GriffinDL84} as vocoder model, as the choice of the vocoder has only limited influence on \gls{asr} recognition \cite{10389782}. 
The encoder gets as input phoneme sequences $a_1^M$ augmented with word separations symbols.\ 
The final text encoder states ${h^{\prime}}_1^M$ are used as input to a duration predictor $f_{\scaleto{DUR}{3pt}}$ which computes $d_m =f_{\scaleto{DUR}{3pt}}({h'}_1^M)$, with $d_i$ being the number of repetitions for each encoder state to up-sample ${h'}_1^M$ to ${h'}_1^T$.\ 
The system incorporates a fixed set of speakers, represented as a lookup table embedding which is used as a local conditioning signal for each coupling block in the flow decoder network.\ 
For data generation, we uniformly draw random speaker labels.
	\vspace{0.5cm}
\section{Experiments}
	\vspace{0.5cm}
\subsection{Data and Language Models}
\label{subsec:data}

Our \gls{asr} models are trained using real 960h LibriSpeech (LBS)\cite{povey2015librispeech}, while the \gls{tts} system utilizes only the clean 460h subset.\ 
We use two additional datasets for the evaluation of the effect of the domain shift:
\begin{itemize}[leftmargin=*]
	\item \textbf{\med{}:} a sub-corpus of the biomedical translation task of the yearly Conference on Machine Translation~(WMT).\ 
	We utilize the English side of the English-German \med{} test dataset of 2022~\cite{neves-EtAl:2022:WMT} as our development set~(dev-22).\ 
	For testing, we utilize \med{} datasets of the 2021~(test-21) and 2023~(test-23) shared tasks.\ 
	For the language model we use the English side of all language pairs in the UFAL Medical Corpus~\cite{ufal}.\ 
	\vspace{1em}
	\item \textbf{\gls{mtg}~\cite{mtgjson}:} an open database\footnote{We accessed the database on January 14th 2025} for the trading card game ``Magic''.\ 
	We use the card and flavor texts as well as the card rules.\ 
	The processed corpus contains 130k lines after de-duplication. We split 1k for both a development and test set.\ 
	The remainder is used for training the language model.
\end{itemize}

\glsreset{lm}
In addition to the official \gls{lbs} \gls{lm}, for each domain data we estimate a 4-gram \gls{lm} via KenLM~\cite{heafield2011kenlm}.\ 
The statistics shown in \cref{tab:ppl_oov} motivates our choice of the data.\ 
While the \med{} text data is characterized by specialized terminology, complex sentence structures, and a formal language, the \gls{mtg} sentences are concise, follow a rule-based syntax.\ 
The shift in the domain is evident in the high perplexity and \gls{oov} rate of the LBS \gls{lm} on both the \med{} and \gls{mtg} development sets.\ 
The higher \gls{oov} rate on \med{} compared to \gls{mtg} also highlights the substantial presence of medical terms.\ 
Moreover, the different levels of linguistic difficulty are reflected in varying perplexities of the UFAL and \gls{mtg} \glspl{lm}, further confirming the simple sentence structure of \gls{mtg} compared to \med{}.\ 
We also train \gls{lstm} \glspl{lm} as the combination with stronger \gls{lm} can further underline the effect of mismatched domain.\ 
The \gls{lm} perplexities are reported in Table \ref{tab:ppl_lstm}.\ 
The experimental results in \cref{tab:tts2} indicate that the effects of domain shift become even more pronounced when combining an \gls{lstm} \gls{lm} to the models with limited or no acoustic label context.\

\begin{table}[t]
	\setlength{\tabcolsep}{0.25em}\renewcommand{\arraystretch}{1.2}  
	\centering \footnotesize 
	\caption{The number of running words and vocabulary size for the LBS, \med{}, and \gls{mtg} datasets. 
		We show the \gls{oov} percentage along with the 4-gram \gls{lm} 
		\gls{ppl} of each task on different development sets.}
	\label{tab:ppl_oov}
	\begin{tabular}{|c||c|c||c|c|c|c|c|c|}
		\hline
		\multirow{2}{*}{\gls{lm}} & Running& Vocab & \multicolumn{2}{c|}{\gls{lbs}} & \multicolumn{2}{c|}{\med{}} & \multicolumn{2}{c|}{\gls{mtg}} \\ \cline{4-9}
		&Words & Size & \gls{oov}{\scriptsize[\%]}& \gls{ppl} & \gls{oov}{\scriptsize[\%]}& \gls{ppl} & \gls{oov}{\scriptsize[\%]} & \gls{ppl}´\\ 
		\hline
		\hline
		\gls{lbs} & 800M&200k& 0.4&\phantom{0}146& 8.7&1508 & 2.7&1048\\\hline
		UFAL & 183M&177k&5.2&2307 & 1.2&\phantom{0}496& \multicolumn{2}{c|}{-} \\ \hline
		\gls{mtg} & \phantom{00}2M&\phantom{0}38k&6.0&1107 & \multicolumn{2}{c|}{-} &0.8 &\phantom{00}34 \\	
		\hline
	\end{tabular} 
\end{table}
\begin{table}[t]
	\setlength{\tabcolsep}{0.25em}\renewcommand{\arraystretch}{1.2}  
	\centering \footnotesize 
	\caption{The word-level perplexity and number of parameters of our \gls{lstm} language models on all tasks for different label units.\ 
	For the \gls{bpe}-level language models we re-normalized the perplexity to word-level. }
	\label{tab:ppl_lstm}
	\begin{tabular}{|l|c|c|c|c|c|c|}
		\hline
		\multirow{2}{*}{Label Unit}& \multicolumn{3}{c|}{\#Parameters} & \multicolumn{3}{c|}{Perplexity}\\ \cline{2-7}
		& \gls{lbs} & \med{} & \gls{mtg} & LBS & \med{} & \gls{mtg}\\ 
		\hline
		\hline
		Word & 244M & 218M & 55M & \phantom{0}78 & 165 & 20.8\\ \hline
		\gls{bpe} 10k & \multicolumn{3}{c|}{25M} & 105 & 302 & 25.2 \\ \hline
		\gls{bpe} 5k & \multicolumn{3}{c|}{19M} & 111 & 334 &  26.0 \\
		\hline
	\end{tabular}
\end{table}

\subsection{Setting}
\label{subsec:setting}
\if\cameraready1
For training we utilize the toolkit RETURNN~\cite{doetsch2017returnn}.\ Decoding of HMM based models use RASR for the core algorithms, and a recent ongoing extension for \gls{ctc} and \gls{trans}  decoding~\cite{rybach2011rasr,zhou2023rasr2}.
\else
For training and decoding of all our systems we utilize two toolkits
\cite{zeyer2018:returnn,wiesler2014rasr}.\
\fi 
For more information on training hyper parameters and decoding settings, we refer to an example of our configuration setups\footnote{\if\cameraready1{\url{ https://github.com/rwth-i6/returnn-experiments/2025-domain-shift}}\else BLIND\fi}.

\subsubsection{ASR systems}
\label{subsubsec:asr-exp}
We used the standard sequence-level cross-entropy criterion for training of \gls{aed} from scratch, and augmented with the sum over all alignments~(full-sum criterion) for \gls{ctc}.\ 
All remaining context-dependent time synchronous models follow a multistage training pipeline~\cite{Zhou+Michel+:2022}.\
We first train a smaller zero-order label context alignment model using full-sum and following the model's label topology: (1) posterior HMM~\cite{Raissi+Zhou+:2023} for \gls{fh} models and (2) \gls{ctc} for the \gls{trans} models.\ 
After a forced alignment, we use the alignment for a fixed path Viterbi training.\ 
Viterbi training using cross-entropy loss has a lower complexity in computation and memory compared to full-sum training which considers all alignment paths following the specific label topology.\ 
In addition, we can make use of efficient sequence chunking techniques. We also found the models to converge faster when using a fixed alignment path for the first part of the training.\ 
We then continue with the regular full-sum loss for the second half of the training.\ 
For the triphone \gls{fh} model without HMM state tying the computation of the state marginals over the phoneme set to the power of three is not feasible.\ An overview of the number of epochs for each model at each stage is shown in \cref{tab:baselines}.\ 
For the phoneme based models we use the official phoneme inventory from the LBS lexicon, by unifying the stressed phonemes and applying end-of-word distinction~\cite{zhou:phoneme-transducer:2021}.\ 
Our \gls{bpe}-based \gls{trans} and \gls{aed} models use a vocabulary of size 5k and 10k, respectively.
All acoustic models use a 12-layers Conformer encoder with an internal dimension of 512~\cite{gulati2020conformer}.\ 
The alignment models use a recurrent encoder consisting of 6 bi-directional \gls{lstm} layers with 512 nodes per direction, having $\sim$46M parameters.\ 
The \gls{aed} model training relies on an auxiliary \gls{ctc} loss.\ 
We use one cycle learning rate schedule~(OCLR) with a peak LR of around 8e-4 over 90\% of the training epochs, followed by a linear decrease to 1e-6~\cite{smith2019super,Zhou+Michel+:2022}.\ 
As optimizer we use  Adam with Nesterov momentum.\ \
We use the standard window of $25$ milliseconds~(ms) with $10$ms shift for feature extraction, resulting in 80 and 40 dimensional log-mel features for \gls{aed} and Gammatone filterbank features~\cite{schluter2007Gammatone} for all other models, respectively.\ 
SpecAugment is applied to all models~\cite{park2019specaugment}.\ 
The \gls{lstm} language models for all tasks and label units consists of 2 \gls{lstm} layers of 1024 hidden dimension trained with initial (LBS: 0.5, UFAL: 2, \gls{mtg}: 50) epochs of constant learning rate of $1.0$ followed by a linear decrease over (LBS: 2.5, UFAL: 10, \gls{mtg}: 50) epochs to 1e-7.\

\subsubsection{Text-To-Speech System}
\label{subsubsec:tts-exp}
For our \gls{tts} model we follow closely the parameters from an existing setup~\cite{kim_glow-tts_2020}, where we increased the hidden dimensions from 192 to 256.\ 
The model is trained for 400 epochs with Adam optimizer and standard OCLR with a peak learning rate of 5e-4.\ 
For the input features, we use 80-dimensional globally normalized log-mel features with a 50ms window and a 12.5ms shift.\ 
The optimized temperature factor $\tau$ of \eqref{eq:tts} is 0.55 for our experiments.

\subsection{Synthetic Test Data Generation}
\label{subsec:datagen}

The \gls{tts} system in our work is used to generate test data that has similar acoustic conditions as the training data but with a different language model domain.\ 
The main constraint is to generate a synthetic LBS development set under one key condition: the \gls{asr} accuracy on the synthesized transcriptions of the original LBS test (dev) data must match that on real \gls{lbs} test (dev) data, as reported in \cref{tab:baselines}.\
For this purpose, we leveraged the capability of Glow-TTS to generate different variants of dev-other via the temperature factor.\ 
Other experimental results relying on non-probabilistic architectures such as FastSpeech-2 \cite{Ren-2020-FastSpeech2Fasta} failed to meet this constraint, and were therefore discarded.\ 
Moreover, since for a given $\tau$ the Gaussian sampling of the latent variable $z_1^T$ in \eqref{eq:tts} is not deterministic, we ensured that the resulting \gls{wer} of our systems remained stable.\ 
Our experimental results confirmed the variance was negligible, with an absolute \gls{wer} variance of just 0.1\% on, e.g., \med{} dev-22 set.\ 
In addition to matching the \gls{wer} itself, we also checked that the ratio of substitutions, insertions and deletions does not differ much between the real and the synthetic \gls{lbs} data.

\begin{table}[t]	
	\setlength{\tabcolsep}{0.25em}\renewcommand{\arraystretch}{1.2}  
	\centering \footnotesize 
	\glsreset{lm}
	\caption{Our baseline models with different label units and label context lengths~(Ctx) trained on real \gls{lbs} 960h and evaluated using 4-gram and \gls{lstm} \glspl{lm} on both real and synthetic LBS data.\
		For each model we report the number of Viterbi~(VIT) and full-sum~(FS) training epochs, as well as the number of model parameters~(PM). }
	\label{tab:baselines}
	\begin{tabular}{|c||c|c||c|c|c||c|c|c||c|c|c||} 
		\hline			
		\multirow{4}{*}{{Model}}& \multicolumn{2}{c||}{{AM}} &\multicolumn{3}{c||}{ Train}&  \multicolumn{3}{c||}{{4-gram \gls{lm}}} & \multicolumn{3}{c||}{{\gls{lstm} \gls{lm}}} \\ \cline{4-12}
		
		&\multicolumn{2}{c||}{Label} & \multicolumn{2}{c|}{\#Epochs}&\multirow{2}{*}{\#PM} & \multicolumn{2}{c|}{{Real}}  &{TTS}& \multicolumn{2}{c|}{{Real}} &{TTS} \\ \cline{2-5} \cline{7-12}
		&\multirow{2}{*}{Unit} & \multirow{2}{*}{Ctx}&\multirow{2}{*}{VIT}&\multirow{2}{*}{FS}&&  {test-} &  \multicolumn{2}{c||}{\multirow{2}{*}{dev-other}} & {test-} &   \multicolumn{2}{c||}{\multirow{2}{*}{dev-other}} \\ 
		&&&&&[M] &other&\multicolumn{2}{c||}{}&other&\multicolumn{2}{c||}{} \\ \hline \hline
		\gls{ctc}&\multirow{4}{*}{Phon}&0&0&100& 74 &6.6&6.2&\textbf{5.9}&5.5&5.1&\textbf{4.8}\\ \cline{1-1} \cline{3-12}
		\multirow{2}{*}{\gls{fh}}&& 2& \multirow{4}{*}{20}&0&76& 6.7 & 6.1 & \textbf{6.0}& 5.4& 4.8& \textbf{4.6}\\ \cline{3-3} \cline{5-12}
		&&\multirow{2}{*}{1}&&\multirow{3}{*}{15}&\multirow{2}{*}{75}& 6.0& 5.6& \textbf{5.9}& 5.1& 4.7& \textbf{4.8}\\ \cline{1-1} \cline{7-12}
		\multirow{2}{*}{\gls{trans}}&&&&&&6.3& 5.8& \textbf{6.0}& 5.2& 4.7&\textbf{4.9}\\  \cline{2-3}\cline{6-12}
		&\multirow{2}{*}{\gls{bpe}}&\multirow{2}{*}{$\infty$}&&&87&6.5&5.9&\textbf{6.0}&5.6&5.0&\textbf{5.0}\\ \cline{1-1} \cline{4-12}
		\gls{aed} &&&0&100&97&\multicolumn{3}{c||}{N/A}&5.0&4.6&\textbf{4.3}\\ 
		\hline
		
	\end{tabular}
\end{table}

\begin{table}[t]	
	\setlength{\tabcolsep}{0.3em}\renewcommand{\arraystretch}{1.3}  
	\centering \footnotesize
	\glsreset{lm}
	\caption{The performance of our models on \gls{lbs} dev-other, as well as dev and test sets of the domain data.\ 
		All evaluation datasets in this table are generated using a \gls{tts} system trained on \lbs{}.\ 
		The acoustic models are also trained on \lbs{} 960h.\ For decoding we utilize the 4-gram \glspl{lm} trained on target domain.}
	\label{tab:tts1}
	\begin{tabular}{|c||c||c|c||c||c|c|c|c|c||} 
		\hline			
		\multirow{2}{*}{\#}&\multirow{2}{*}{Model}&\multicolumn{2}{c||}{AM Label}& {LBS}&\multicolumn{3}{c|}{\med{}}& \multicolumn{2}{c||}{\gls{mtg}} \\ \cline{3-10}
		&&\multirow{1}{*}{Unit} & \multirow{1}{*}{Ctx}&  dev-other & dev-22& test-21 & test-23 & dev& test  \\ \hline \hline
		
		1&\gls{ctc}&\multirow{4}{*}{Phon}&0&\textbf{5.9}&10.8& 11.5 &12.2 & 5.4&5.4 \\ \cline{1-2} \cline{4-10}
		
		2&\multirow{2}{*}{\gls{fh}} &&2&6.0&\textbf{10.7}&\textbf{11.1}& 12.0&\textbf{5.2}&\textbf{5.1} \\ \cline{1-1} \cline{4-10}
		
		3&&&\multirow{2}{*}{1}& \textbf{5.9}&10.9&11.5&\textbf{11.9}&5.3&5.4\\ \cline{1-1} \cline{2-2} \cline{5-10}
		
		4&\multirow{2}{*}{\gls{trans}}&&&6.0&11.8 &12.3&12.6 &5.6& 5.2\\  \cline{1-1}  \cline{3-4} \cline{5-10}
		
		5&&\multirow{1}{*}{\gls{bpe}}&\multirow{1}{*}{$\infty$}&6.0&16.9&16.2&16.1&7.9&7.3\\ \cline{1-2} \cline{5-10}		
		\hline
	\end{tabular}
\end{table}

\begin{table}[t]	
	\setlength{\tabcolsep}{0.3em}\renewcommand{\arraystretch}{1.3}  
	\centering \footnotesize
	\glsreset{lm}
	\caption{Similar experiments as in \cref{tab:tts1} with \gls{lstm} \glspl{lm} trained on target domain.}
	\label{tab:tts2}
	\begin{tabular}{|c||c||c|c||c|c|c|c|c|c||} 
		\hline			
		\multirow{2}{*}{\#}& \multirow{2}{*}{Model} &\multicolumn{2}{c||}{AM Label}& {LBS}&\multicolumn{3}{c|}{\med{}}& \multicolumn{2}{c||}{\gls{mtg}}\\ \cline{3-10}
		&&\multirow{1}{*}{Unit} & \multirow{1}{*}{Ctx}&  dev-other & dev-22& test-21 & test-23 & dev& test \\ \hline \hline
		
		1&\gls{ctc}&\multirow{4}{*}{Phon}&0& 4.8&9.5&9.5&10.1&4.8&4.6\\ \cline{1-2} \cline{4-10}
		
		2&\multirow{2}{*}{\gls{fh}} &&2& 4.6&\textbf{8.5}&\textbf{9.2}&\textbf{9.4}&\textbf{4.6}& \textbf{4.3}\\ \cline{1-1} \cline{4-10}
		
		3&&&\multirow{2}{*}{1}& 4.8&9.3&9.8&9.8&4.9& 4.7\\ \cline{1-1} \cline{2-2} \cline{5-10}
		
		4&\multirow{2}{*}{\gls{trans}}&&&4.9&9.6&10.3&10.3&4.9& 4.5\\  \cline{1-1}  \cline{3-4} \cline{5-10}
		
		5&&\multirow{2}{*}{\gls{bpe}}&\multirow{2}{*}{$\infty$}&5.0&15.6&14.7&13.8&6.4&6.0\\ \cline{1-2} \cline{5-10}
		
		6&\gls{aed} &&&\textbf{4.3}&12.1&12.0&12.1&5.0&4.9\\ 
		\hline
	\end{tabular}
\end{table}

\begin{table}[t]
	\setlength{\tabcolsep}{0.25em}\renewcommand{\arraystretch}{1.2}  
	\centering \footnotesize 
	\caption{The comparison between closed and open vocabulary decoding using word and \gls{bpe} level \glspl{lm}, respectively. }
	\label{tab:openvocab}
	\begin{tabular}{|c|c|c||c|c|c||c|c|}		
		\hline
		\multirow{2}{*}{Model}& \multicolumn{2}{c||}{ Decode}&  \multicolumn{3}{c||}{\med{}}& \multicolumn{2}{c|}{\gls{mtg}} \\ \cline{2-8}
		&\gls{lm}& Vocab& dev-22 & test-21 & test23 & dev& test \\
		\hline
		\hline
		\multirow{2}{*}{\gls{trans}} & Word &Closed & 15.6 & 14.7& 13.8& 6.4& 6.0\\  \cline{2-8}
		& \multirow{2}{*}{\gls{bpe}}& \multirow{2}{*}{Open}& 14.3&14.7& 14.5& 6.7 & 6.3\\  \cline{1-1} \cline{4-8}
		\gls{aed} &&&12.1&12.0&12.1&5.0&4.9\\
		\hline
	\end{tabular}
\end{table}

\subsection{Comparative Analysis of Domain Shift}
\label{subsec:comparative}

 In addition to an experimental design for \gls{asr} and \gls{tts} systems that guarantees comparable \gls{asr} accuracy on the synthetic dev-other across all models, we also made deliberate choices to build highly optimized systems that are mutually comparable.\ 
However, this approach did not eliminate some of the differences, outlined in \cref{subsubsec:asr-exp}, that persist mainly due to specific requirements of each model.\ 
 For instance, the diphone \gls{fh} trained without relying on Gaussian mixture model alignment or HMM state tying, closely relates to the phoneme-based \gls{trans} with first-order label context.\ Both models can be seen as context-dependent alternatives to the phoneme-based \gls{ctc}, making all three representatives of models that allow for lexical prefix tree decoding with differences in label topology and model definition~\cite{raissiinvestigating}.\ 
The triphone \gls{fh} also falls into this category.\ 
However, by extending label context-dependency to the right, the model is able to distinguish between the left-center phonemes appearing in different right context.\ These alignment states are tied in a diphone topology.\

In \cref{tab:baselines}, we present the \gls{asr} WER of our models on real \gls{lbs} test data, as well as both real and synthetic \gls{lbs} dev data.\ 
It is possible to see that with both 4-gram and \gls{lstm} \glspl{lm} all models obtain comparable results, with a slight advantage for \gls{aed} with neural \gls{lm}.

 Following the four-fold comparative scenario discussed in \cref{sec:expdes}, we present our domain shift results using the 4-gram and \gls{lstm} \gls{lm}s in \cref{tab:tts1,tab:tts2}, respectively.\ In the first comparison between the diphone \gls{fh} and the phoneme based \gls{trans} models, the small performance gap observed with the count-based LM is nearly closed when a stronger LM is used.\ However, the modeling of the co-articulation effect via right label context in triphone \gls{fh} (Line 2) stands out particularly outperforming all other approaches when using a \gls{lstm} \gls{lm}.\ 
 All mentioned phoneme-based models are theoretically more robust to domain shift due to their limited label context-dependency and shorter label unit.\ 
In our third experimental comparison between phoneme based and \gls{bpe} based transducers, we confirm the widely observed effect that BPE-based models with full label context tend to exhibit a stronger \gls{ilm}.\ A common motivation for using \gls{bpe} based models is their support for open-vocabulary decoding, which offers greater flexibility in generating out-of-domain words.\ In contrast, closed vocabulary decoding is restricted by the pronunciation lexicon but may potentially benefit from language model and vocabulary adaptation.\ To provide direct comparison, we compare open and closed vocabulary decoding in \cref{tab:openvocab}.\ It is possible to see that there is no significant gain from the open vocabulary decoding for the \gls{bpe} based transducer model with infinite context.\ 
One possible explanation is that the small amount of target domain data for training the \gls{bpe} based \gls{lm} in combination with a strong \gls{ilm} learned on the source domain limits the freedom of the model for coming up with unseen words.\
Concerning the fourth comparison, while the phoneme \gls{ctc} shows slightly greater robustness compared to the diphone \gls{fh} and phoneme \gls{trans}, its performance lags behind that of the triphone \gls{fh}.\ Moreover, we observe that the \gls{bpe} \gls{aed} with infinite context also exhibits substantial performance degradation under domain mismatch.

\section{Discussions}
Given that our acoustic models never see text data from target domain, we assume that the performance differences are heavily conditioned by the type of \gls{ilm} each model learns.\ This can extend also to the hybrid HMM models when using a powerful encoder such as Conformer.\ A detailed experimental comparison of \glspl{ilm} is beyond the scope of this work and left for future research.\ 
A limitation of \gls{bpe} based models could be the change in segmentation.\ 
While on LibriSpeech the \gls{bpe} token to word ratio for the \gls{aed} model is~1.1, it becomes~1.6 for Medline and~1.3 for \gls{mtg}, meaning that there is a strong change in the distribution of labels.\ 
In that case, it could be preferable to use smaller \gls{bpe} segmentations which might not be optimal for LibriSpeech, but perform better under domain shift.\ 
Moreover, since the common \gls{tts} systems in the literature use phonemes as label unit, an investigation on possible \gls{bpe} based \gls{tts} can offer further understanding.

\section{Conclusion}

Our proposed analysis lays the foundation for a principled comparison of domain shift behavior for different \gls{asr} architectures.\ 
By generating synthetic data with acoustic conditions similar to those in the training data, we showed the effect of the domain shift for popular \gls{seq2seq} models with different label units and frame-level label posteriors.\ We showed that robustness to domain shift during decoding is not necessarily determined by the choice of the decoder architecture, but rather specific modeling choices.\ While making no claim as to exhaustiveness, we have shown that the choice of label units, context length, and topology plays a significant role for domain shift behavior across different architectures. This insight goes beyond the usual simplified dichotomy between classic hybrid HMM and novel \gls{seq2seq} systems.\ At this stage of our research, we conclude that the phoneme based models with limited context present more robustness under domain shift when incorporating an external language model trained on the target domain irrespective of the underlying ASR architecture.\ 

\section{Acknowledgements}

This work was partially supported by NeuroSys, which as part of the
initiative “Clusters4Future” is funded by the Federal Ministry of
Education and Research BMBF (03ZU2106DA). We appreciate Benedikt Hilmes' enthusiasm for playing Magic The Gathering and pitching the idea of trading card games as a limited vocabulary domain. We would like to thank Simon Berger, Mohammad Zeineldeen, and Atanas Gruev for providing the Phoneme Transducer, BPE Attention, and BPE Transducer baselines, respectively.

\bibliographystyle{IEEEtran}
\bibliography{mybib}

\end{document}